\documentclass[twocolumn,english,aps,prd, floatfix,amssymb,superscriptaddress,a4paper]{revtex4-1}
\usepackage{lipsum}
\usepackage{amsmath}

\usepackage{amssymb}
\usepackage{graphicx}
\usepackage{physics}
\usepackage{slashed}
\usepackage{hyperref}
\usepackage[dvipsnames]{xcolor}
\usepackage{float}
\usepackage[caption=false]{subfig}

\begin{document}

\title{{Echoes and defects in the Calogero model} }  

\author{Benjamin Liégeois}
\author{Ramasubramanian Chitra}
\author{Nicolò Defenu}
\affiliation
{Institute for Theoretical Physics, ETH Zurich, 8093 Zurich, Switzerland}




\date{\today}

 \begin{abstract}
%
In this work, we extend the study of the interplay between scaling symmetries and statistics to one-dimensional fluids by studying the Calogero model in a harmonic trap modulated through time. The latter harbors an interpretation in terms of free particles imbued with exclusion statistics and is an example of a scale invariant fluid in one-dimension displaying SO(2,1) dynamical symmetry preserved by harmonic traps. Taking advantage of the dynamical symmetry, two experimentally relevant drive protocols spanning both quasi-static and non-adiabatic regimes are investigated and universal signatures of the interactions and exclusion statistics are uncovered in the ground-state echo amplitude and closely related ground-state fidelity. In particular, under both periodic modulation and slow drive through the gapless point of the trap frequency, enhanced interactions and exclusion are shown to favor the proliferation of defects and to hinder their annihilation, which leads to a universal decrease of ground-state fidelities and echo amplitudes.  We also show that increasing exclusion sparks a sharp suppression of the likelihood of intermediate echoes beyond those imposed by the commensurability of a periodic drive and the natural frequency of the trap. %
\end{abstract}
\vspace{-4mm}
\maketitle
\vspace{-4mm}


\section{Introduction}

 Symmetries are fundamental to our understanding of physical systems. The  associated existence of conserved quantities greatly simplifies the study of their dynamics. More generally,  a system might harbor dynamical symmetries (also referred to as hidden or spectrum-generating symmetries)  resulting in additional dynamical invariants. In particular, scale invariant fluids form a notable class of systems that displays rich dynamical symmetries.  A defining feature of such fluids is their behavior under a dilation $x \to \zeta x$, for which the Hamiltonian scales as  $H(\zeta x) = \zeta^\alpha H(x)$ where  $\alpha$ is some real exponent. Such scaling has dramatic consequences: for example, it leads to exceptionally simple thermodynamic properties such as equations of state depending solely on the ratio of temperature to chemical potential as opposed to both independently~\cite{DalibardExperiment}. Remarkably, even when confined by a parabolic potential, scale invariant fluids may still display $\mathrm{SO(2,1)}$ hidden symmetry, again with deep consequences both for the time evolution and equilibrium properties of the fluid.

Experimental observations of effectively scale invariant fluids are ubiquitous, notably thanks to ultracold atomic gas platforms. This includes a myriad of systems among which the unitary
Fermi gas in three dimensions~\cite{UnitaryFermi}, the two-dimensional (2D) Bose gas~\cite{SU11_HiddenSymmetry} in the weakly interacting limit and
2D Fermi gas away from the crossover regime~\cite{hofmann2012quantum,PuneetDefenuScience}, the Tonks-Girardeau gas in one dimension~\cite{TonksGirardeau} as well as all systems effectively described through a Gross-Pitaevskii equation in all dimensions~\cite{GrossPitaevskii}. 

In this context, the $\mathrm{SO(2,1)}$ dynamical symmetry~\cite{BreathersEchoes} was shown to be responsible for the emergence of breathers and echoes~\cite{ZhouCitation} in 2D Bose gas experiments~\cite{DalibardExperiment}. These studies focused on the dynamics of  $\mathrm{SO(2,1)}$-symmetric systems prepared in the ground-state of their non-interacting counterparts (i.e. the ground-state of the trapping Hamiltonian). The crucial presence of a $\mathrm{SO(2,1)}$ dynamical symmetry equally enables the study of the dynamics initialized in the fully interacting ground-state, which is the focus of the present work. 

Beyond symmetry, another crucial aspect of quantum mechanics is the notion of statistics. Particle statistics have dramatic consequences for the phenomenology of a system. For example, while fermions cannot occupy the same state due to Pauli repulsion, bosons may macroscopically condense into a coherent state leading to quantum coherent phenomena such as superfluidity and superconductivity. Statistics also influence dynamical properties: for example,  it was predicted that dynamical crossings of infinitely degenerate quantum critical points (encountered in effective bosonic theories) are non-adiabatic~\cite{Graf_InfDegenerate,defenu2018dynamical} independently on the drive's rate or functional form~\cite{Nicolo_AdiabaticBreakdown}.  In stark contrast, when crossing quantum critical points in effective fermionic theories, the corrections to adiabaticity vanish in the slow drive limit through a non-analytic scaling, which is predicted by the Kibble-Zurek mechanism~\cite{KZM, KZM_Fermions_I, KZM_Fermions_II, KZM_Fermions_III, KZM_Fermions_IV}. 

Lower dimensional many-body systems offer the potential for generalized notions of statistics, beyond the bosonic and fermionic limits, raising the interest for the exploration of the interplay between scaling symmetries and statistics in this setting.  In this work, we consider one dimension,  where a different notion of statistics is often considered: exclusion statistics~\cite{HaldaneStatistics}. The Calogero model~\cite{Calogero, Sutherland_I, Sutherland_II, Moser} is a paradigmatic one-dimensional model displaying both a scaling symmetry and an interpretation in terms of free particles imbued with exclusion statistics~\cite{Calogero_Exlusion}.  This model displays the dynamical $\mathrm{SO(2,1)}$ symmetry alluded to above, making it an ideal  system to explore the interplay between  exclusion statistics and hidden symmetries.
In this work, we  investigate  the dynamics of the Calogero model in a time-dependent harmonic trap, which explicitly reduces the conformal invariance to the $\mathrm{SO(2,1)}$ spectrum generating symmetry.  We study both periodic modulations of the trap frequency and slow drives through the gapless point.  Enhanced exclusion and interactions are predicted to favor the proliferation of defects and to hinder their annihilation. Our claims are then verified through the computation of ground-state fidelities and echo amplitudes, in which stronger exclusion and interactions are shown to universally lower the chance of revivals. 

\section{Model, symmetries, and exclusion interpretation}

The Calogero model~\cite{Calogero, Sutherland_I, Sutherland_II, Moser} describes $N$ identical particles in one dimension interacting pairwise via an inverse-square potential and confined by a harmonic well
\begin{equation}\label{Hamitonian}
{H}=-\frac{1}{2} \sum_{i=1}^N \frac{\partial^2}{\partial x_i^2}+\sum_{i<j} \frac{\lambda
(\lambda-1)}{\left(x_i-x_j\right)^2}+\frac{1}{2} \omega^2 \sum_{i=1}^N x_i^2.  
\end{equation}

The eigenfunctions $\Psi_n(\left\{x_i\right\})$ of the Calogero model are known exactly~\cite{EigenstatesCalogero}. In particular, the ground-state takes the form
\begin{equation}\label{GroundStateWavefunction}
\Psi_{0}\left(\left\{x_i\right\};\omega\right)=\mathcal{N}_{N, \lambda} \omega^{\tfrac{N}{4}[1+\lambda(N-1)] } e^{-\frac{1}{2} \omega \sum_i x_i^2} \Delta\left(\left\{x_i\right\}\right)^{\lambda},
\end{equation}
where $\Delta\left(\left\{x_i\right\}\right) \equiv \prod_{j<k}\left(x_j-x_k\right)$ is the Vandermonde determinant and $\mathcal{N}_{N, \lambda}$ a known normalization constant. The associated energy spectrum is given by~\cite{Calogero}
\begin{equation}\label{Spectrum}
    E_n=\frac{N}{2} \omega+\lambda \frac{N(N-1)}{2} \omega+ \sum_{i=1}^N n_i\omega,
\end{equation}
where the  excitation numbers $n_i$ sum to $n$ and obey the constraint: $n_i \leq n_{i+1}$.

\begin{figure}[h!]
    \centering
    \includegraphics[width=0.48\textwidth]{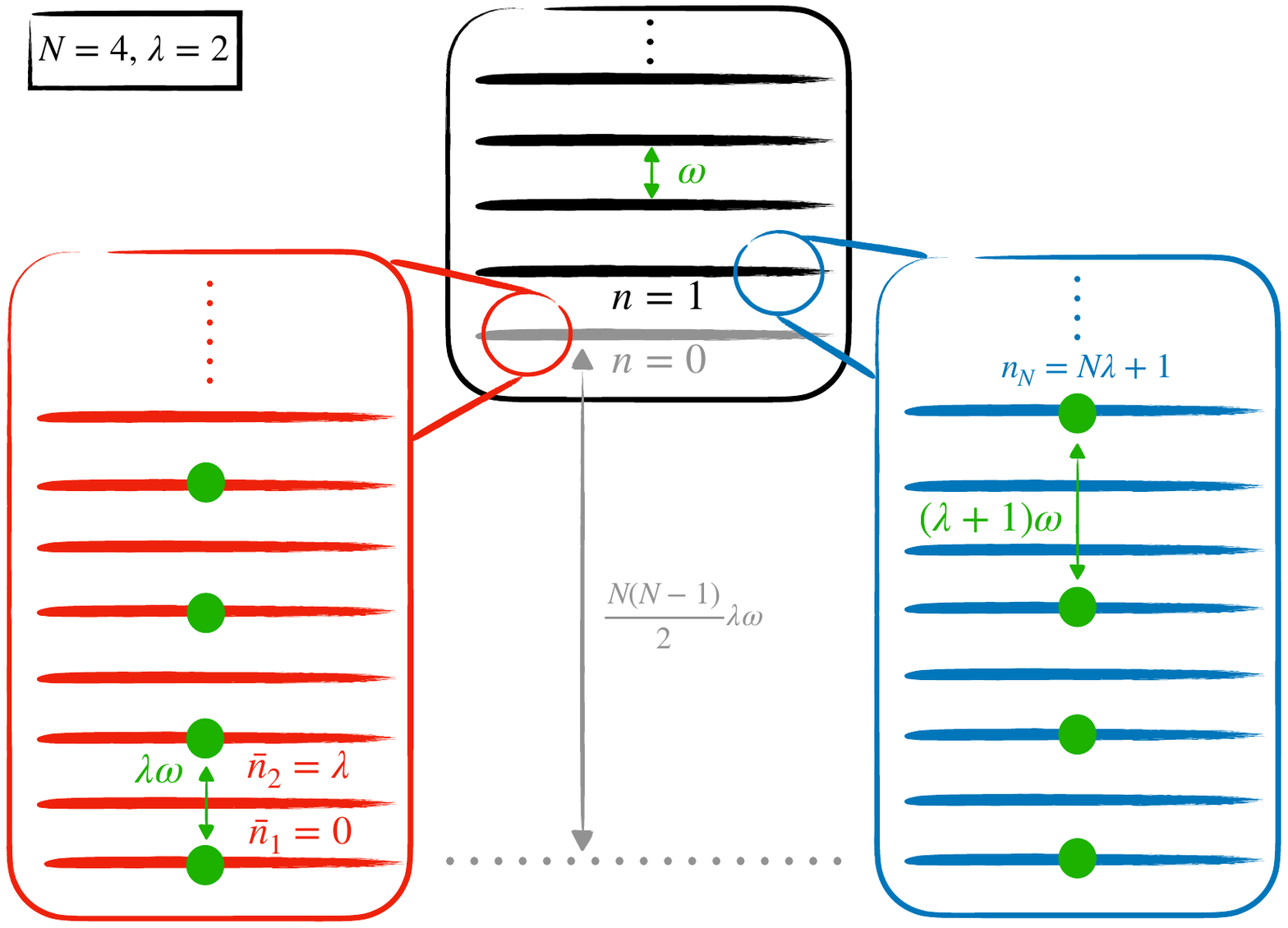}\vspace{-1mm}
    \caption{Schematic representation of the exclusion statistics interpretation of the Calogero model. The central black box represents the evenly spaced many-body spectrum with an offset by $N(N-1)\lambda\omega/2$ and each state is labeled by $n$. Each such many-body state $n$ can be viewed as resulting from the filling of a single-particle spectrum without offset with $N$ particles labeled $i$ together with the constraint that the single-particle level $\bar{n}_i$ occupied by the $i$th particle is at least separated by $\lambda$ from the level of the previous particle $\bar{n}_{i-1}$. An illustration is given for $\lambda = 2$ and $N=4$. The many-body ground-state in gray in the central black box corresponds to the configuration minimizing the numbers $\bar{n}_i$ as shown in the red box on the left. The first excited state corresponds to a configuration which increases these numbers by one unit as shown in the blue box on the right, and so on for higher excited states. \vspace{-2mm}  }
    \label{fig:Exclusion}
\end{figure}

To see exclusion statistics  directly, one first  defines pseudo-excitation numbers as $\bar{n}_i=n_i+(i-1) \lambda$~\cite{PhysMathCalogeroPolychronakos}. The expression of the spectrum in terms of these numbers becomes identical to that of a system of free harmonically-trapped bosons with an additional constraint on the quantum numbers
\begin{equation}
\bar{n}_i \leq \bar{n}_{i+1}-\lambda.
\end{equation}
This can be interpreted as an exclusion principle that requires the pseudo-excitation numbers to be at least $\lambda$ apart, which manifests a generalized statistics interpretation~\cite{Calogero_Exlusion}. Particular cases include the free boson gas at $\lambda=0$ and the Tonks-Girardeau gas (hard-core bosons) at $\lambda=1$, while other values describe  one dimensional  Haldane anyons~\cite{HaldaneStatistics}.  Note that the numbers $\bar{n}_i$ are not necessarily integers, but they are modified through integer increments. The minimal allowed nonnegative set of values for $\bar{n}_i$ obeying the selection rules determines the ground-state uniquely while excited states are built through all integer increments thereof. Figure~\ref{fig:Exclusion} displays a schematic representation of this exclusion statistics interpretation. 

Notably, the scaling of the kinetic energy and pair-wise interaction in Eq.~(\ref{Hamitonian}) are identical under dilation, making the Calogero Hamiltonian in the absence of a trap scale-invariant. As pointed out earlier and as we will show in the following, the presence of a harmonic trapping potential breaks scale invariance but preserves the $\mathrm{SO(2,1)}$ dynamical symmetry~\cite{SU11_HiddenSymmetry},  independently of whether the trapping potential is modulated through time, i.e. when $\omega=\omega(t)$. This allows us to go beyond static properties of the Calogero model and to uncover more general features of its dynamics associated with its $\mathrm{SO(2,1)}$ hidden symmetry. A direct consequence of the dynamical symmetry is the predictable dynamics of the time-dependent many-body wavefunction $\psi_n(t)$ as the harmonic frequency is varied in time $\omega(t)$. We indeed have that if the system occupies the eigenstate $\Psi_n$ with trapping potential $\omega(t_0)=\bar{\omega}$ at any instant $t_{0}$, the time-evolved state at time $t$ with trapping frequency $\omega(t)$ can be obtained through the formula~\cite{AdolfoDerivationDensity, ScalingDynamics}
\begin{equation}\label{ScalingEvolution}
\psi_n\left(\left\{x_i\right\}, t\right)=\frac{\Psi_n\left(\left\{\frac{x_i}{\xi(t)}\right\}\right)  e^{i \frac{\dot{\xi}(t)}{2 \xi(t)} \sum_{i=1}^N x_i^2-i E_n \tau(t)}}{\xi(t)^{\frac{N}{2}}}  ,
\end{equation}
with $\tau(t)=\int_{t_{0}}^t \frac{d t^{\prime}}{ \xi^2\left(t^{\prime}\right)}$ and $E_n$ the energy of the eigenstate $\Psi_n$ chosen as the initial state.  The scaling factor $\xi(t)>0$ is the solution of the Ermakov-Milne differential equation 
\begin{equation}\label{Ermakov}\ddot{\xi}(t)+\omega(t)^{2} \xi(t)=\frac{\bar{\omega}^2}{\xi^{3}(t)},\end{equation}
with $\xi(t_{0})=1$ and $\dot \xi(t_{0})=0$.

We now show how scale invariance can be argued to lead to a $\mathrm{SO(2,1)}$ dynamical symmetry  even in the presence of a time-dependent harmonic trap, and sketch how scaling symmetry leads to the scaling dynamics described by Eq.~(\ref{ScalingEvolution}). Let us assume a Hamiltonian of the form ${H}(t)=H_0+\omega^2(t)H_{\mathrm{trap}}$, with $H_0=K+V$ where $K=-\tfrac{1}{2}\sum_i \partial_{i}^2$ is the standard non-relativistic kinetic Hamiltonian,  $H_{\mathrm{trap}}=\tfrac{1}{2}\sum_i x_i^2$ the unit-frequency harmonic trap and the potential $V$ scales like the kinetic part $K$ under rescaling $V(\{\zeta x_i\}) = \zeta^{-2} V(\{ x_i\})$. This form encompasses in particular the driven Calogero model. As shown in Ref.~\cite{ScalingDynamics}, taking Eq.~(\ref{ScalingEvolution}) as an ansatz for the dynamical state, imposing $i\dot{\psi}= H(t)\psi$ and using the assumed scaling of the potential $V$ straightforwardly results in $\xi$ satisfying the Ermakov equation~(\ref{Ermakov}). The occurrence of these scaling dynamics is intimately linked to the fact that $V$ scales like $K$ under rescaling, or in other words that the untrapped Hamiltonian $H_0$ is scale invariant. A special property of the harmonic potential  $H_{\mathrm{trap}}$ is the fact that its commutator  with any scale-invariant base Hamitonian yields the generator of scaling transformations: $[H_{\mathrm{trap}}, H_0] = iQ$ where $Q=\tfrac{1}{2}\sum_j \left[(-i\partial_{j})  x_j+x_j  (-i\partial_{j})\right]$. Consequently, a closed $\mathrm{SO(2,1)}$ algebra can be generated out of linear combinations of $H_{\mathrm{trap}}, H_0$ and $Q$. Defining $T_1=\frac{1}{2}\left(H_0-H_{\mathrm{trap}}\right)$,  $T_2=\frac{Q}{2}$ and $T_3=\frac{1}{2 }\left(H_0+H_{\mathrm{trap}}\right)$, we indeed find that $\left[T_1, T_2\right]=-i T_3, ~~\left[T_2, T_3\right]=i T_1, ~~\left[T_3, T_1\right]=i T_2$. The time-dependent Hamiltonian can then be explicitly expressed in terms of these generators as $ H(t)  = [1-\omega(t)^2] T_1 + [1+\omega(t)^2] T_3$. The associated evolution operator can subsequently be factorized into a product of exponentials of the generators $T_i$~\cite{su11disentangle, su11_Decomposition}. The "spectrum-generating" action of these generators transforming eigenstates into one another is known and thereby formally solves the full time evolution, resulting in Eq.~(\ref{ScalingEvolution}). 

All in all, we have argued that given a solution of the Ermakov-Milne equation for a particular functional form of the modulated frequency $\omega(t)$, Eq.~(\ref{ScalingEvolution}) provides the exact evolution of the state of the system when prepared in one of its eigenstates. In this paper, we study the system prepared in its ground-state $\Psi_0$. Particularizing Eq.~(\ref{ScalingEvolution}) to $n=0$ and using the expression for $\Psi_0$ and $E_0$ following from Eqs.~(\ref{GroundStateWavefunction}) and~(\ref{Spectrum}), we straightforwardly find an expression for the evolved ground-state of the Calogero model $\psi_0(t)$ in terms of the solution $\xi(t)$ of the Ermakov-Milne equation~\cite{ExactCoherentSutherland_AdolfoCite}
\begin{equation}\label{GroundStateDynamics}
\frac{\psi_0(t)}{\mathcal{N}_{N, \lambda}}=e^{\frac{-\Omega(t)}{2} \sum_i x_i^2-iE_0\tau(t)}\Delta\left(\left\{x_i\right\}\right)^{\lambda}\hspace{-1mm}\left(\frac{\bar{\omega}}{\xi^2(t)}\right)^{\tfrac{N}{4}[1+\lambda(N-1)]},
\end{equation}
with $\Omega(t)=-i \frac{\dot{\xi}(t)}{\xi(t)}+\frac{\bar{\omega}}{\xi^2(t)}$ and $\mathcal{N}_{N, \lambda}$ the normalization constant referred to in Eq.~(\ref{GroundStateWavefunction}). This exact result provides the basis for the computation of observable quantities of interest. We present such observables and expose our expectations regarding their dynamics in the following.

\section{Defect proliferation out of equilibrium}\label{sec:expectations}

\begin{figure}[h!]
    \centering
    \includegraphics[width=0.46\textwidth]{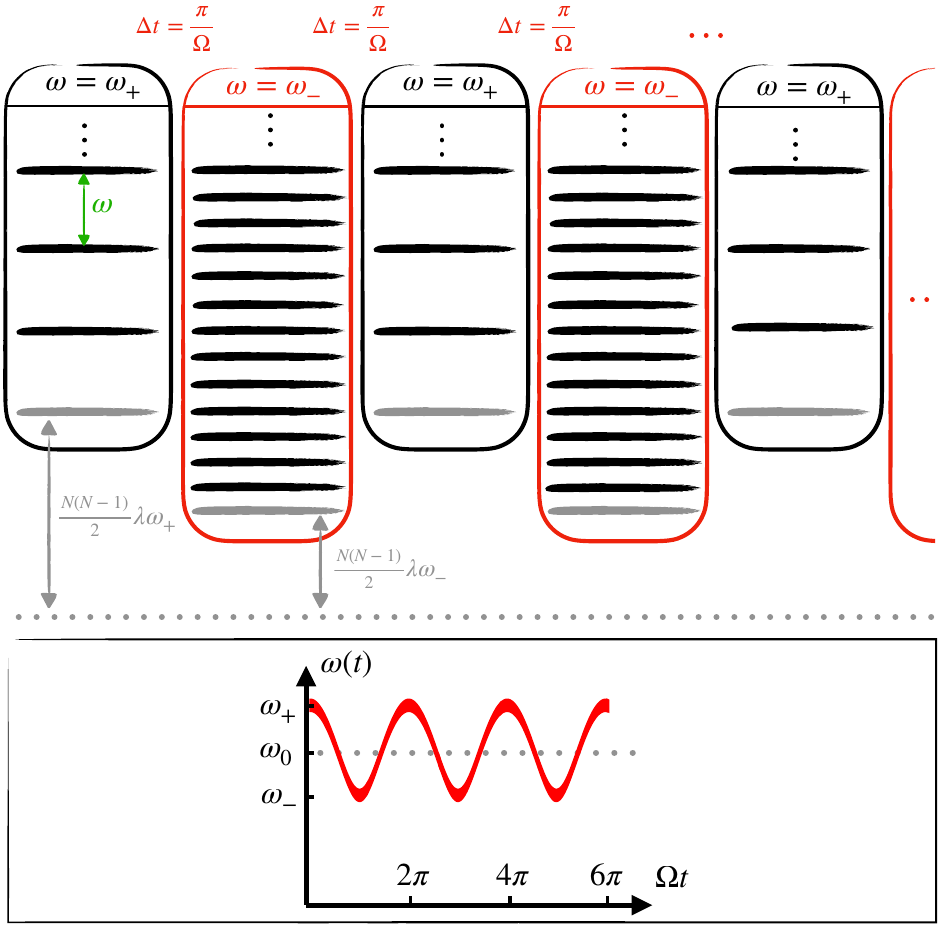}
    \vspace{-4mm}
    \caption{Schematic representation of the protocol corresponding to an adiabatic drive through the gapless point. Starting with the system in its ground-state (in gray in the leftmost black box) trapped with a given frequency $\omega= \omega_+$, the first protocol continuously drives the frequency $\omega$ through $0$ (third box in blue with full level degeneracy) before restoring the trapping frequency $\omega= \omega_+$ to its original value.  } 
    \label{fig:Drive}
\end{figure}
 
 We first present a heuristic argument for how statistics affect defect proliferation when the harmonic trap frequency $\omega$ is varied in time. Figures~\ref{fig:Drive} and~\ref{fig:Drive2}, which depict specific choices of drive protocols for $\omega$ and their effects on the many-body spectrum, are both useful in illustrating the analysis.

 Note that when $\omega$ is lowered,  the energy level spacing decreases and  the spectrum has a downward shift in energy proportional to the interaction strength $\lambda$.  Starting in the ground-state $E_0$ at some fixed $\omega$, as  $\omega$ decreases,  the number of  energy levels with higher quantum number $n$ that cross (below) the initial energy level $E_0$  grows with  $\lambda$. 
 Large values of $\lambda$ are therefore expected to favor transitions to higher quantum numbers when the trapping potential is weakened. Interactions thus facilitate the proliferation of defects. Conversely, when $\omega$ increases, the spectrum is shifted up and the level spacing increases. The energy of the populated levels with highest quantum numbers increases fast with respect to the energy of the ground-state, which renders the annihilation of defects more difficult.

The physical picture provided above essentially stems from  an exclusion constraint on free particles. In the single-particle energy spectrum picture, there is no longer a $\lambda$-dependent shift but rather the constraint for the quantum numbers $\bar{n}_i$ to be at least $\lambda$ apart. As the spectrum is filled from the bottom with a fixed number of particles $N$, the single-particle energies occupied are more spread out and larger for increasing $\lambda$. When the level spacing is lowered through a modulation of $\omega$, the increased density of states and the greater initial span in energy of particles lower the importance of the constraints.  This offers more reconfiguration possibilities into single-particle states with higher quantum numbers $\bar{n}_i$. As the level spacing is restored, not only are the defects to be annihilated at higher levels, but the constraints become more restricting and the number of relaxation paths from these higher states to the ground-state is reduced.

In our analysis, we focus on two closely related observables, relevant in cold atom experiments. In particular, we wish to investigate in both cases the deviation of the dynamical state (harboring defects) with respect to a given choice of reference state. For quasi-static drives, we focus on the ground-state fidelity $f(t)$ with the instantaneous adiabatic ground-state, while in non-adiabatic settings, as the initial state is more relevant we consider the closely related echo amplitude $e(t)$. These are defined as 
\begin{equation}
    f(t)=\left|\braket{\psi^{ad}_{0}(t)}{\psi_{0}(t)}\right|^2, \; e(t)=\left|\braket{\psi_{0}(0)}{\psi_{0}(t)}\right|^2,
\end{equation} 
where $\psi^{ad}_{0}(t)$ is the instantaneous equilibrium ground-state, $\psi_{0}(0)$ is the initial ground-state and $\psi_0({t})$ is the time-evolved dynamical state.  Based on the heuristic discussion presented earlier, we expect the enhanced proliferation of defects sparked by higher values of $\lambda$ to induce a decrease in the ground-state echo amplitude and fidelity. 

\section{Validation through experimentally relevant protocols}

\begin{figure}[h!]
    \centering
    \includegraphics[width=0.44\textwidth]{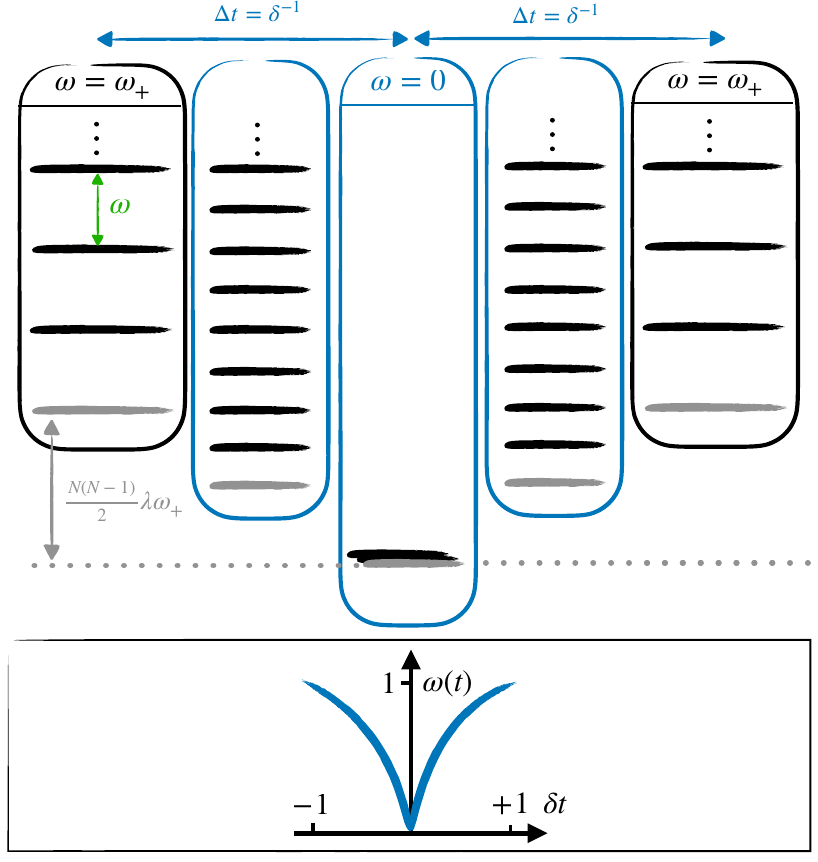}
    \vspace{-4mm}
    \caption{Schematic representation of the Floquet experimental protocol investigated in this work. Starting with the system in its ground-state (in gray in the leftmost black box) trapped with a given frequency $\omega= \omega_+$, the protocol drives the trapping frequency to a lower value $\omega_-$ before being restored, this procedure being repeated with a period $2\pi/\Omega$.  } 
    \label{fig:Drive2} 
\end{figure}

In the following, we study two experimentally-relevant protocols beyond the sudden and linear quenches studied in the literature~\cite{QuenchCalogeroDunkle, delCampo_ExactQuantumDecay, DelCampoCite}.  To explore the inherently out-of-equilibrium and non-adiabatic regimes, we consider the Floquet protocol \begin{equation}\label{FloquetDrive}
     \omega^2(t)\equiv \omega^2_0[1+h\cos(\Omega t)], 
 \end{equation}  
for $t\in [0,\infty)$ where the gas undergoes periodic cycles of compression and expansion. To address adiabatic regimes closer to equilibrium, a drive through the gapless point $\omega=0$ is investigated (akin to a sweep through the critical gap closing point discussed in the introduction)
\begin{equation} \label{KibbleZurekDrive}
    \omega^2(t)\equiv |\delta t|^{2 z \nu},
\end{equation}
with rate $\delta$ and $t\in [-1/\delta,1/\delta]$. In other words, the gas is slowly let free of its trap before being trapped again. The quantity $z\nu$ in Eq.~(\ref{KibbleZurekDrive}) is {\it a priori} simply a parameter describing the dynamical protocol. The model may, however, be viewed as an effective model for a many-body system ramped across its quantum critical point in the spirit of Refs.~\cite{defenu2018dynamical, BreakdownAdiabatic, Graf_InfDegenerate, Nicolo_AdiabaticBreakdown}. Within this perspective the exponent $z\nu$ may represent the dynamical critical exponent for the gap scaling. The point at which the trap is zero corresponds to the gapless critical point in this system, where the energy levels are degenerate. This specific functional form of the protocol is of particular interest because of its universality in the adiabatic limit: any drive through $\omega=0$, provided it is slow enough, can be expanded to lowest order yielding the form given by Eq.~(\ref{KibbleZurekDrive}). We, therefore, expect the response of the system to this protocol to encompass the response of all drives which can be expanded into a power law in the adiabatic limit. The protocols and their effects on the many-body spectrum are illustrated respectively in Figs.~\ref{fig:Drive} and~\ref{fig:Drive2}. 

\section{Results}  

The Ermakov-Milne equation~(\ref{Ermakov}) can be solved for both protocols: in terms of non-linear combinations of Mathieu functions for the Floquet drive in Eqs.\,\eqref{FloquetDrive} and generalized Airy functions for the slow drive protocol in Eq.\eqref{KibbleZurekDrive}. The details of the calculations in the case of the Floquet drive are reproduced in Appendix~\ref{Appendix:ErmakovFloquet}, while the corresponding computations for the drive through the gapless point closely follow those obtained in Ref.~\cite{Nicolo_AdiabaticBreakdown}. Given Eq.~(\ref{GroundStateDynamics}), this yields an exact expression for the evolved ground-state. As detailed in Appendix~\ref{Appendix:FidelityEcho} in agreement with Ref.~\cite{delCampo_ExactQuantumDecay}, the ground-state fidelity can be derived in terms of the solution to the Ermakov-Milne equation and reads
 \begin{equation}\label{FidelityErmakov}
     f(t)=\left(\frac{4\omega(t)\bar{\omega}}{\alpha(t) \xi^2(t)}\right)^{N[1+\lambda(N-1)]/2},
 \end{equation} 
where the function $\alpha(t)$ is given by 
\begin{equation}\label{alphaFunction}
\alpha(t)=\left(\omega(t)+\frac{\bar{\omega}}{\xi^2(t)}\right)^2+\left(\frac{\dot{\xi}(t)}{\xi(t)}\right)^2.\end{equation}
An analogous expression is found for the echo amplitude $e(t)$ and can be obtained by replacing $\omega(t)$ by $\bar{\omega}$ in Eqs.~(\ref{FidelityErmakov}) and~(\ref{alphaFunction})
\begin{equation}\label{EchoeAmplitudeErmakov}
     e(t)=\left(\frac{4\bar{\omega}^2}{\tilde{\alpha}(t) \xi^2(t)}\right)^{N[1+\lambda(N-1)]/2},
 \end{equation} 
where $\tilde{\alpha}(t)$ is given by 
\begin{equation}\label{alphaFunction}
\tilde{\alpha}(t)=\left(\bar{\omega}+\frac{\bar{\omega}}{\xi^2(t)}\right)^2+\left(\frac{\dot{\xi}(t)}{\xi(t)}\right)^2.\end{equation}

Note that the interaction and statistics parameter $\lambda$  appears explicitly in Eq.~(\ref{FidelityErmakov}).  Making the $\lambda$-dependence explicit $f_\lambda(t)$, one observes that the fidelity (and closely-related echo amplitudes) are given by the corresponding free bosonic equivalent $f_0(t)$ raised to a $\lambda$-dependent exponent 
\begin{equation}\label{ExpDecay}
    f_\lambda(t) = [f_{0}(t)]^{\lambda \tfrac{N(N-1)}{2}}.
\end{equation}
In other words, the more these quantities are driven away from unity, the faster their exponential decay with $\lambda$. As $\lambda$ fixes the exclusion statistics, our results clearly show that the fidelity and echo amplitude are inherently dependent on the statistics of the particles. 

We first address the case of the Floquet protocol. This study is of great interest given that the underlying $\mathrm{SO(2,1)}$ symmetry structure of the Calogero Hamiltonian is intimately linked to the emergence of parametric resonance.  It is well known that parametric instabilities can emerge in such systems when both the amplitude $h$ and the modulation frequency $\Omega$ are varied.  In the $(h, \tfrac{\Omega}{\omega_0})$ space,  parametric resonance delineates regions which are stable from those which are unstable giving rise to a pattern of stability lobes (known as Arnold tongues~\cite{ArnoldTongues}). The stability~\cite{FloquetGritsev} and periodicity~\cite{BreathersEchoes} properties of the system's response to the drive~(\ref{FloquetDrive}) are universal features associated with the $\mathrm{SO(2,1)}$ dynamical symmetry. In particular, the occurrence of particular beat structures and parametric instabilities is uniquely determined by the characteristic exponent of the Mathieu equation associated with the drive~(\ref{FloquetDrive}), which depends only on $h$ and the ratio $\omega_0/\Omega$. Note that the parametric instability corresponds to the points in $(h, \omega_0/\Omega)$-space where the Mathieu exponent becomes complex.

\begin{figure}[h!]
    \centering
    \includegraphics[scale=0.4]{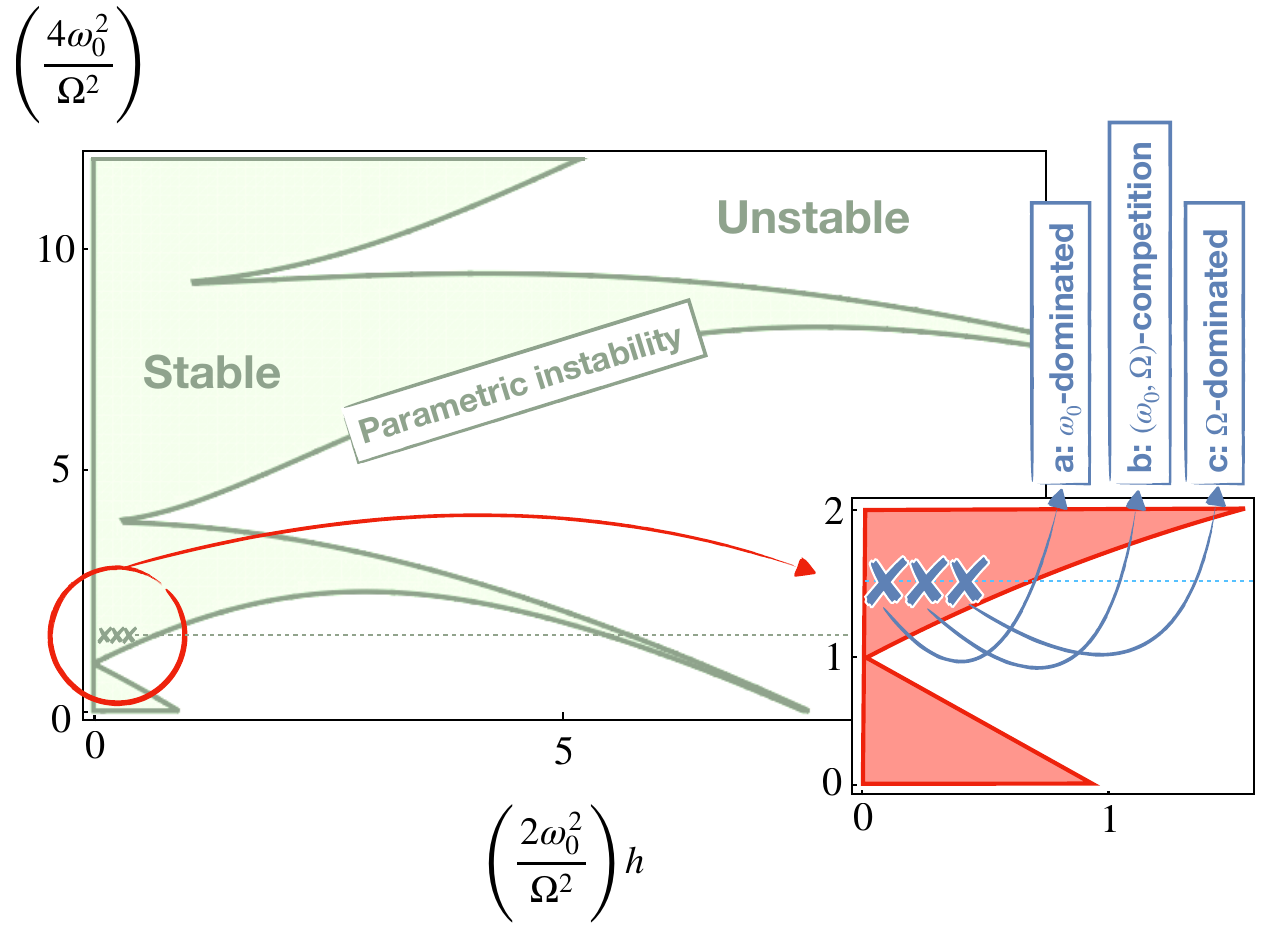}
    \vspace{-3mm}
    \caption{ Stability diagram   of periodically driven systems with $\mathrm{SO(2,1)}$ dynamical symmetry. Inset: Zoom-in on the parametric values considered in Fig.~\ref{fig:EchoAmplitudeStable}. These correspond to regimes where the echo amplitude is (a) dominated by the natural trap frequency $\omega_0$ for low drive amplitudes $h$, (b) a result of the competition between the driving frequency of the trap $\Omega$ and $\omega_0$ for intermediate drive amplitudes $h$, and (c)  dominated by the driving frequency of the trap $\Omega$ for drive amplitudes $h$ approaching the parametric instability. 
    } 
    \label{fig:MathieuStability}
\end{figure}

Perfect ground-state revivals, however, are directly linked to the commensurability of the two frequencies $\omega_0$ and $\Omega$.  For all $\lambda$ and $N$, these perfect revivals occur at  times which are solely determined  by  the parameters $(h, \omega_0/\Omega)$. These times are shared identically by the classical Mathieu oscillator, parametric quantum harmonic oscillator (corresponding to $N=1,\lambda=0$) and the corresponding periodically driven trapped free Bose ($N>1, \lambda=0$) and Tonks-Girardeau gas ($N>1, \lambda=1$), with no effect of interactions $\lambda$. Given the $\lambda$-independence  of these features, the periodically driven Calgero model (which generalizes the latter models to $N>1, \lambda>0$) will display analog beat structures, perfect echoes and instabilities dictated uniquely and solely by parameters of the drive $(h, \omega_0/\Omega)$. The universal stability diagram~\cite{FloquetGritsev} associated with periodically driven systems with $\mathrm{SO(2,1)}$ dynamical symmetry (including the Calogero model and the other systems mentioned above) is displayed in Fig.~\ref{fig:MathieuStability}. 

While the occurrence of perfect echoes and the beat structures are solely dictated by $(h, \omega_0/\Omega)$, we will see that the echo amplitude in between exact revival points is greatly influenced by the interactions $\lambda$, consistently with Eq.~(\ref{ExpDecay}). In other words, $\lambda$ crucially determines the likeliness of echoes beyond those ensured by commensurability of the drive and the natural trapping frequency. This reveals an intricate and strong interdependence between interactions and statistics and echoes in the Floquet setting. 
 
\begin{figure}[h!]
    \centering
    \includegraphics[scale=0.4]{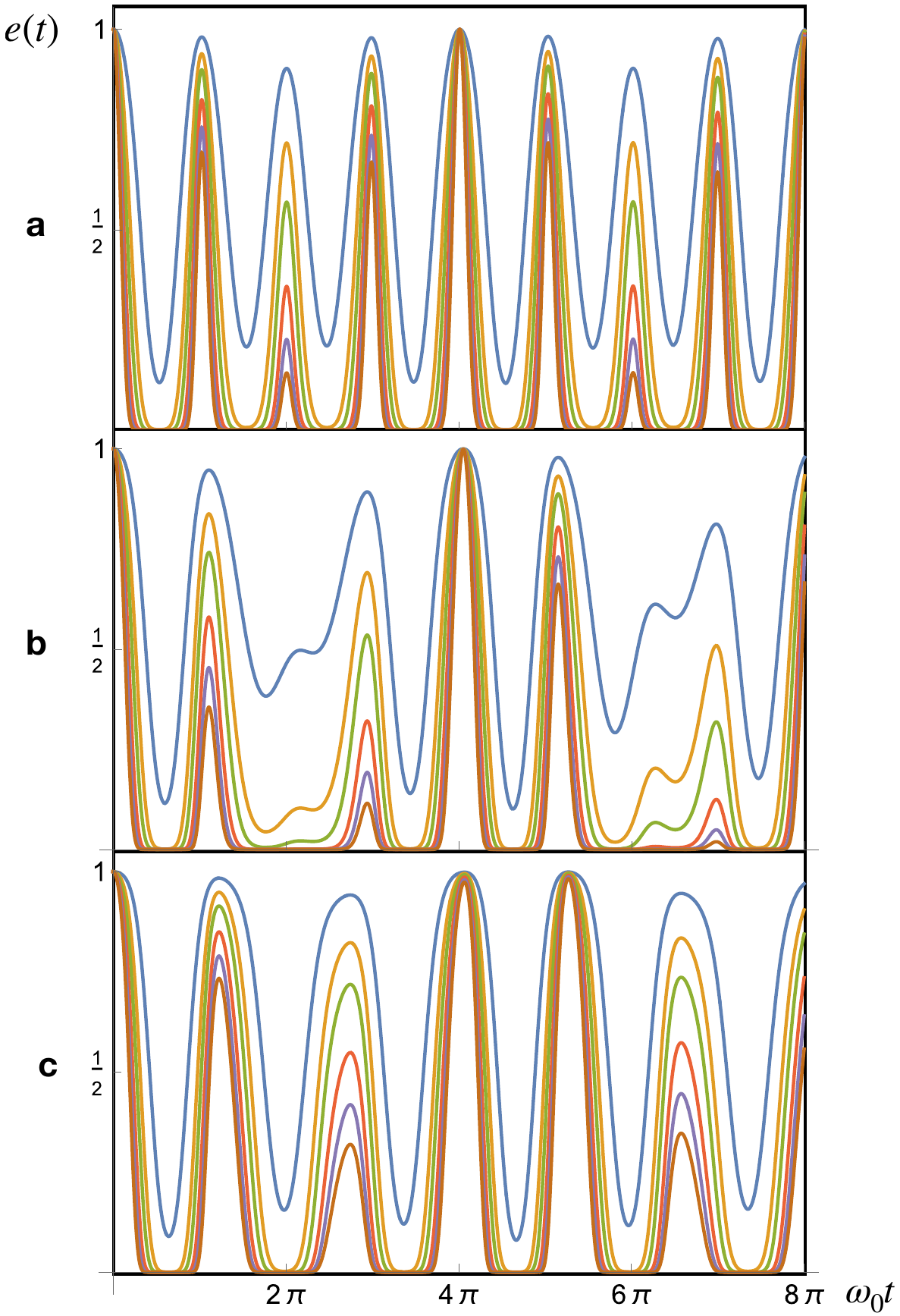}
    \vspace{-3mm}
    \caption{ground-state echo amplitude $e(t)$ for a periodic drive with a non-resonant but commensurate drive frequency  
$\Omega = 3/2$ in units of the trap frequency $\omega_0$, for $N=10$. Each subpanel displays $e(t)$ for an amplitude of the drive $h$ given by (a) 1/8 (b) 1/3, and (c) 1/2. The curves in each panel from top to bottom correspond to $\lambda = 0,\tfrac{1}{4},\tfrac{1}{2},1,\tfrac{3}{2}, 2$. The points considered in parameter space of the drive are referenced within the stability digram in Fig.~\ref{fig:MathieuStability}.} 
    \label{fig:EchoAmplitudeStable}
\end{figure}

Our results for the ground-state echo amplitude $e(t)$ of the Calogero model in the parametrically stable regimes of the Floquet drive are displayed in Fig.~\ref{fig:EchoAmplitudeStable} for fixed $N$ and several interaction strengths $\lambda$, and this for three sets of parameters $(\tfrac{\omega_0}{\Omega}, h)$ in the stable region with different degrees of stability. The position of the corresponding regimes in parameter space are indicated in the stability diagram displayed in Fig.~\ref{fig:MathieuStability}. 

Given commensurate frequencies ${2\omega_0}$ and ${\Omega}$, one observes that at times where the two corresponding sinusoidal signals [say $\cos(2\omega_0)$ and $\cos(\Omega)$] are in phase, an echo occurs with absolute certainty: the corresponding amplitude is unity and is unaffected by $\lambda$. In contrast, at times where the former are not in phase, the echo amplitude is strongly suppressed by larger values of $\lambda$ through the exponential suppression embodying Eq.~(\ref{ExpDecay}). This is most visible in the intermediate regime of the drive's amplitude $h$, when both the natural and drive frequencies compete. Here, one witnesses a dramatic suppression of the intermediate peaks occurring between the commensurability-imposed perfect echoes [see Fig.~\ref{fig:EchoAmplitudeStable}(b)]. In the  small $h$ regime, the echo amplitude is dominated by the natural frequency of the trap $2\omega_0$ as seen in Fig.~\ref{fig:EchoAmplitudeStable}(a), while for large values of $h$ (approaching the parametric instability), the echo amplitude is dominated by the frequency of the drive $\Omega$ as seen in Fig.~\ref{fig:EchoAmplitudeStable}(c). The effective frequency structure in intermediate regimes such as those of Fig.~\ref{fig:EchoAmplitudeStable}(b) is dictated by the Mathieu exponent, which depends only on the parameters of the drive (and not $\lambda$), and interpolates between those two cases. In the limiting cases of vanishing $h$ or amplitudes $h$ approaching the instability, intermediate echoes would become ever more likely and the resulting perfect echoes would occur with frequencies $2\omega_0$ and $\Omega$ respectively.  Figs.~\ref{fig:EchoAmplitudeStable}(a) and~\ref{fig:EchoAmplitudeStable}(c) however clearly demonstrate that small deviations from these limits already spark large drops in the likelihood of intermediate echoes.  To summarize, we observe under Floquet drives a universal decrease in the chances of intermediate revivals through interactions and exclusion (beyond those perfect echoes imposed by commensurability). 
 \begin{figure}[h!]
    \centering
    \includegraphics[scale=0.4]{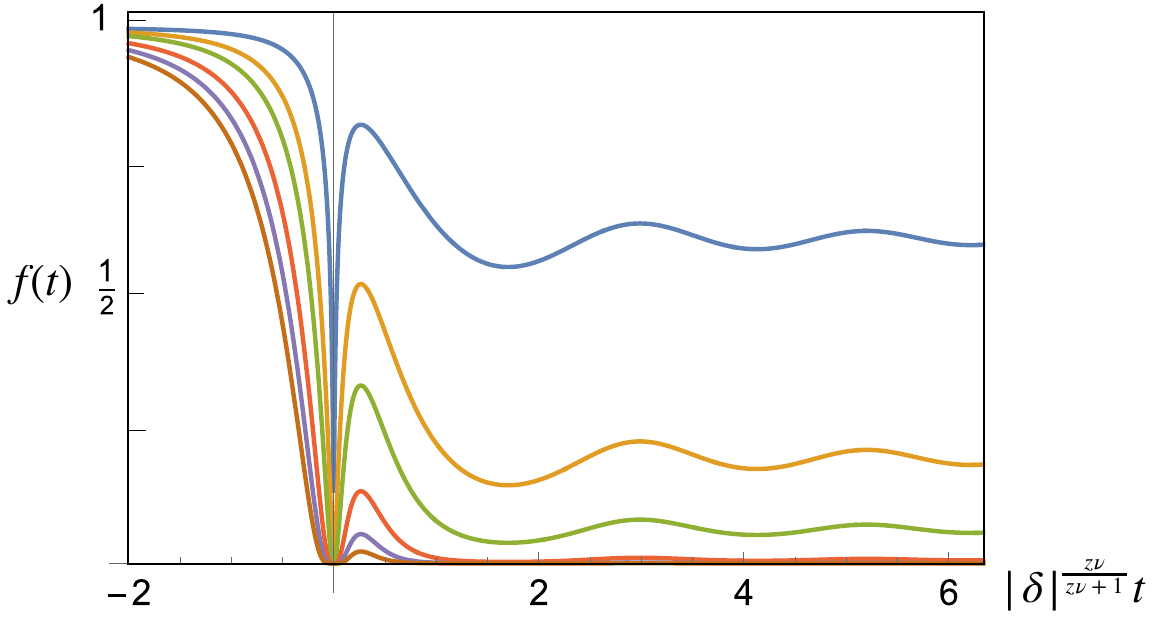} 
    \vspace{-3mm}
    \caption{ground-state fidelity $f(t)$ for a slow quench through $\omega=0$ with parameters $z\nu\equiv 1/4, \; N=10 $ and from top to bottom $\lambda \in \{0,\tfrac{1}{4},\tfrac{1}{2},1,\tfrac{3}{2}, 2\}$.}
    \label{fig:Fidelity}
\end{figure}
 
In the slow drive limit, the effects of particle statistics are substantially more uniform. Indeed, in Fig.~\ref{fig:Fidelity} we observe that in the asymptotic $\delta\to 0$ limit, the fidelity is significantly lowered with respect to its initial (unit) value (even in the bosonic case). This is due to the proliferation of defects. The  asymptotic behavior can be obtained exactly for the slow quench through $\omega=0$ as 
\begin{equation}\label{Asymptotics}
\lim _{t \rightarrow \infty} f(t)=[\sin (p \pi)]^{N(1+\lambda(N-1))},\end{equation} with $p=(2+2 z \nu)^{-1}$, showing that the asymptotic fidelity of a harmonic quantum Calogero gas driven across its critical point is always a constant irrespectively of the rate exponent $z\nu$ of the ramp time dependence. While this constant depends on $z\nu$ as well as $N$ and $\lambda$, it is universal in its independence from the rate $\delta$. This is consistent with the universal breakdown of of adiabaticity predicted to follow the crossing of infinitely degenerate quantum critical points (encountered in effective free bosonic theories)~\cite{Nicolo_AdiabaticBreakdown, Graf_InfDegenerate}. Our results, however, extend those predictions to interacting settings or equivalently to non-trivial exclusion statistics: the asympotic values~(\ref{Asymptotics}) display a clear dependence on $\lambda$. As seen in Fig.~\ref{fig:Fidelity}, interactions (or exclusion statistics) $\lambda$ induce a further drop in the asymptotic fidelity. This is consistent with the enhanced proliferation and hindered annihilation of defects through interactions predicted in Sec.~\ref{sec:expectations}.

To sum up,  while interactions result in a sizable quantitative drop in the fidelity for slow drives, they spark qualitative features such as the sharp suppression of the intermediate echoes in the  Floquet setting.  

\section{Discussion}

In summary, we investigated the interplay between exclusion statistics and spectrum generating symmetries in the harmonically confined Calogero model out of equilibrium. Our studies considered both periodic and adiabatic modulations of the trap potential, focusing on the ground-state echo amplitudes and fidelity. Both quantities are dramatically suppressed  with increasing exclusion quantified by the  interaction strength $\lambda$. Given a periodic drive and a commensurate natural trap frequency, we predict a sharp suppression of the likelihood of intermediate echoes (between those imposed by commensurability) in the presence of interactions. Additionally, our study of the effects of interactions forecasts a substantial drop in the asymptotic ground-state fidelity reached following a slow drive through $\omega=0$, independent of the driving rate $\delta$. All of these findings are consistent with favored proliferation and hindered annihilation of defects sparked by interactions, which can be interpreted as a result of exclusion statistics.

These conclusions are underpinned by the intricate and fascinating interplay between statistics and dynamical symmetries of scale-invariant fluids embodied by the Calogero model, which we hope to verify in near-term future experiments. 

Although $1/r^2$ interactions have not yet been the subject of targeted experiments, a possible realization of the Calogero model in cold Bose atom systems with dipole-dipole interactions was devised in Ref.~\cite{YueYu}. Meanwhile one also could potentially engineer a many-body system out of Efimov states~\cite{Macek2007}, which are bound in an effective $1/r^2$ potential or exploit the tunability of interactions offered through the use of dipolar molecular gases or mediation via fermions or guided photons~\cite{Tune1,Tune2,Tune3}. A direct observation of defects in the fashion of Ref.~\cite{Ko2019} would then allow for experimental observation of the proliferation of defects with enhanced statistics predicted in our work. An even cleverer approach would be to construct and study in a similar fashion SO(2,1) symmetric systems tailored to model interactions already realized in the existing experimental platforms. 

Following this study, it is also of great interest to look into further observables such as the pair-momentum distribution.  Although perhaps less easily amenable to analytical treatment, the latter is highly relevant in cold atom experiments and would provide great additional insights regarding the consequences of statistics and scaling dynamics on correlations and spatial coherence. Further work of great significance includes the investigation of dynamics of the Calogero model breaking the $\mathrm{SO(2,1)}$ dynamical symmetry through modulation through time of the interaction parameter $\lambda$ into a regime where quantum anomalies and bound states may play a role or the study of multispecies and spinful generalizations of the Calogero model. 
Our work also motivates the exploration of the interplay between fractional statistics and dynamics in two dimensional systems like quantum Hall systems hosting anions.

\acknowledgments

This work was supported by an ETH Zürich Research Grant, the Swiss
National Science Foundation (SNSF) under project funding ID: 200021\_207537, by the Swiss Secretariat for Education,
Research and Innovation (SERI) and by the Deutsche Forschungsgemeinschaft (DFG, German Research Foundation) under
Germany's Excellence Strategy EXC2181/1-390900948 (the Heidelberg STRUCTURES Excellence Cluster).

%

\appendix

\section{Derivation of ground-state fidelity and echo amplitudes} \label{Appendix:FidelityEcho} 

Given the dynamical state's evolution~(\ref{GroundStateDynamics}), we now derive expressions for observable quantities of particular interest. In particular, we first focus on the ground-state echo amplitude and fidelity, given respectively by the overlap of the dynamical state (evolved from the initial state taken as the ground-state) and the instantaneous equilibrium (or adiabatic) ground-state. The latter is given by

\begin{equation}\label{AdiabaticGroundState}
\frac{\psi_{0}^{ad}\left(\left\{x_i\right\},t\right)}{\mathcal{N}_{N, \lambda}}= \omega(t)^{\tfrac{N}{4}[1+\lambda(N-1)] } e^{-\frac{1}{2} \omega(t) \sum_i x_i^2} \Delta\left(\left\{x_i\right\}\right)^{\lambda}.
\end{equation}
This is obtained simply by simply replacing $\omega$ in the expression of the static ground-state of the Calgero model~(\ref{GroundStateWavefunction}) by its dynamic equivalent $\omega(t)$. The ground-state fidelity then refers to 
\begin{equation}\begin{aligned}
    f(t) &= \left|\braket{\psi^{ad}_{0}(t)}{\psi_{0}(t)}\right|^2\\ &= \left|\int \prod_{i=1}^N \mathrm{~d}x_i {\psi^{ad}_{0}(\{x_i\},t)}{\psi_{0}(\{x_i\},t)}\right|^2, 
\end{aligned}
\end{equation}
where the implied domain of integration is ${\mathbb{R}^N}$. By inserting the expressions~(\ref{GroundStateDynamics}) and~(\ref{AdiabaticGroundState}) above for the dynamical and adiabatic ground-states, we find that
\begin{equation}\begin{aligned}
f(t)&=\mathcal{N}_{N, \lambda}^{-2}\left(\frac{\omega(t)\bar{\omega}}{\xi(t)^2}\right)^{N[1+\lambda(N-1)]/2}\times\\
& \quad \quad\left|\int \prod_{i=1}^N \mathrm{~d} x_i \,\mathrm{e}^{-\frac{x_i^2}{2}\left(\omega(t)+\Omega(t)\right)} \prod_{j>i}\left| x_i-x_j\right|^{2 \lambda}\right|^2 
\end{aligned}
\end{equation}
with $\Omega(t)=-i \frac{\dot{\xi}(t)}{\xi(t)}+\frac{\bar{\omega}}{\xi^2(t)}$. To obtain a closed form for the expression above, we make use of Mehta's integral~\cite{MethaIntegral}
\begin{equation}
\begin{aligned}
    \mathcal{I}_{\text{Metha}}&\equiv \int \cdots \int \prod_{i=1}^N e^{-x_i^2 / 2} \prod_{1 \leq i<j \leq n}\left|x_i-x_j\right|^{2 \lambda} d x_1 \cdots d x_N\\
 &= (2 \pi)^{N / 2} \prod_{j=0}^{N-1} \frac{\Gamma[1+(j+1) \lambda]}{\Gamma(1+\lambda)}  = \frac{\mathcal{N}_{N, \lambda} }{2^{-\frac{N}{2}[1+\lambda(N-1)]}}.
\end{aligned}
\end{equation}  

\noindent In particular, the  relation
\begin{equation}\begin{aligned}
\left|\int \prod_{i=1}^N \mathrm{~d} x_i \,\mathrm{e}^{-\frac{x_i^2}{2}\left(\omega(t)+\Omega(t)\right)} \prod_{j>i}\left| x_i-x_j\right|^{2 \lambda}\right|^2\\ = \frac{(\mathcal{I}_{\text{Metha}})^2}{|\omega(t)+\Omega(t)|^{N[1+\lambda N(N-1)]}}     
\end{aligned} \end{equation} 
is found by rescaling $x_i$ by a factor $|\omega(t)+\Omega(t)|$. This respectively yields factors $ |\omega(t)+\Omega(t)|^{N}$  from the measure and $|[\omega(t)+\Omega(t)]|^{\lambda N(N-1)}$ from the Jastrow factor. Inserting this result and the value of Metha's integral, the following closed expression for the ground-state fidelity is obtained
\begin{equation}
f(t) =\left(\frac{4\omega(t)\bar{\omega}}{\alpha(t) \xi^2(t)}\right)^{N[1+\lambda(N-1)]/2},
\end{equation}
where the function $\alpha(t)$ is given by 
\begin{equation}
\alpha(t)=\left[\left(\omega(t)+\frac{\bar{\omega}}{\xi(t)^2}\right)^2+\left(\frac{\dot{\xi}(t)}{\xi(t)}\right)^2\right].
\end{equation}

In the case of a periodic drive, we are more interested in the overlap with the initial state. Hence we have to compute

\begin{equation}
e(t) = \left|\braket{\Psi_{0}}{\psi_{0}(t)}\right|^2 = \left|\braket{\psi_{0}(t_0)}{\psi_{0}(t)}\right|^2 \end{equation}
\noindent  which is obtained by replacing $\omega(t)$ by $\bar{\omega}=\omega(t_0)$ in the expression of the overlap with the adiabatic equilibrium ground-state

\begin{equation}
e(t)=\left(\frac{4\bar{\omega}^2}{\tilde{\alpha}(t) \xi^2(t)}\right)^{N[1+\lambda(N-1)]/2},
\end{equation}
where the function $\tilde{\alpha}(t)$ is given by 
\begin{equation}
\tilde{\alpha}(t)=\left[\left(\bar{\omega}+\frac{\bar{\omega}}{\xi(t)^2}\right)^2+\left(\frac{\dot{\xi}(t)}{\xi(t)}\right)^2\right].\end{equation}









\section{Solution of Ermakov-Milne equation for the Floquet drive} \label{Appendix:ErmakovFloquet}

We wish to solve the Ermakov equation 
\begin{equation}
\ddot{\xi}(t)+\omega(t)^2 \xi(t)=\frac{\bar{\omega}^2}{ \xi^3(t)} 
\end{equation}
for particular functional forms of the drive of the frequency $\omega(t)$. The general solution of the Ermakov is given by
\begin{equation}
\xi(t)^2=\left[a x_1(t)+b_1 x_2(t)\right]^2+b_2^2 x_2(t)^2,
\end{equation}
with $b_2=\frac{1}{2a\mathrm{Wr}(x_1,x_2)} $ is given in terms of independent solutions $x_i(t)$ of the corresponding Hill's equation
\begin{equation} {\ddot{x}}_i(t) + \omega^2(t) x_i(t) = 0.\end{equation}

\noindent For the periodic  drive
    \begin{equation} \omega^2(t)\equiv \omega^2_0[1+h\cos(\Omega t)], \end{equation}
we set $t_0=0$ and have $\omega(0) = \bar{\omega} = \omega_0\sqrt{1+h}$ and the corresponding Hill's equation is a Mathieu equation, which is solved through the sine- and cosine-elliptic Mathieu functions 
\begin{equation}\mathcal{M}^S_{\left(\frac{ 2{\omega_0}}{\Omega}\right)^2}\left[\tfrac{1}{2}\Omega t; \left(\frac{2 h \omega^2_0}{\Omega^2}\right)\right], \quad \mathcal{M}^C_{\left(\frac{4 \omega^2_0}{\Omega^2}\right)}\left[\tfrac{1}{2}\Omega t; \left(\frac{2 h \omega^2_0}{\Omega^2}\right)\right].\end{equation}
Mathieu's equation is symmetric under complex conjugation. The real and imaginary parts of solutions hence also qualify as solutions. The corresponding Ermakov solution $\xi(t)$ being a real-valued function, one can express it in terms of the real parts of Mathieu functions (solutions of the Mathieu equations).  In particular, we want to normalize the solutions so that their Wronskian is unity. Omitting the dependence on parameters other than time,  we have already 

\begin{equation}  \dot{\mathcal{M}}^C(0) = 0, \quad {\mathcal{M}}^S(0) = 0.\end{equation}

\noindent We can pick the solutions $x_1$, $x_2$ to be such that $x_1\propto \mathrm{Re}[\mathcal{M}^C]$ and $x_2\propto \mathrm{Re}[\mathcal{M}^S]$. To ensure the unity of the Wronskian, we can further impose 

\begin{equation} {x}_1(0)=1, \quad \dot{x}_2(0) =1, \end{equation}

\noindent so that 

\begin{equation} \begin{aligned}
    \mathrm{Wr}(x_1,x_2)&\equiv x_1(t)\dot{x}_2(t)-\dot{x}_1(t) x_2(t)\\ &= x_1(0)\dot{x}_2(0)-\dot{x}_1(0) x_2(0)\\ &=1, 
\end{aligned}\end{equation}
where we use the stationarity of the Wronskian for linearly independent solutions of linear differential equations. We can pick 
\begin{equation}  {x}_1(t)=\frac{\mathrm{Re}\left[\mathcal{M}^C(\tfrac{1}{2}\Omega t)\right]}{\mathrm{Re}\left[\mathcal{M}^C(0)\right]}, \quad {x}_2(t)=\frac{2}{\Omega}\frac{\mathrm{Re}\left[\mathcal{M}^S(\tfrac{1}{2}\Omega t)\right]}{\mathrm{Re}\left[\dot{\mathcal{M}}^S(0)\right]}.  \end{equation}
\noindent The solution of the Ermakov equation can then in general be written as
\begin{equation}
\xi^2(t)=\left[a x_1(t)+b x_2(t)\right]^2+\frac{1}{4 a^2} x_2(t)^2,   
\end{equation}
where we used $b_2=[{2a\mathrm{Wr}(x_1,x_2)}]^{-1} = \frac{1}{2a} $. The boundary conditions $\xi(0)=1$ and $\dot{\xi}(0)=0$ follow from the stationarity of the initial state (taken at time $t_0=0$) and impose 
\begin{equation}\xi^2(0) = a^2 x_1^2(0) =a^2 \equiv 1, \end{equation}
which yields $a=1$. Since 
\begin{equation} \dot{(\xi^2)}(0) = 2\xi(0)\dot{\xi}(0) = 2 \dot{\xi}(0), \end{equation}
we can impose $\dot{(\xi^2)}(0)\equiv 0$. We thus have 
\begin{equation}\begin{aligned}
\dot{(\xi^2)}(0) &= 2 \left[ a x_1(0)+ bx_2(0) \right][a \dot{x}_1(0)+b \dot{x}_2(0)](\tfrac{\Omega}{2})\\
&+ \frac{1}{4a^2}[2 x_2(0)] \dot{x}_2(0)(\tfrac{\Omega}{2}) = ab\Omega \equiv 0,  
\end{aligned}\end{equation}
which yields $b=0$.  The solution of the Ermakov-Milne equation follows
        \begin{equation}
            \xi^2(t)= x_1^2(t)+\frac{1}{4} x_2^2(t),
        \end{equation} 
with

\begin{equation}{x}_1(t)=\frac{\mathrm{Re}\left[\mathcal{M}^C(\tfrac{1}{2}\Omega t)\right]}{\mathrm{Re}\left[\mathcal{M}^C(0)\right]}, \quad {x}_2(t)=\frac{2}{\Omega}\frac{\mathrm{Re}\left[\mathcal{M}^S(\tfrac{1}{2}\Omega t)\right]}{\mathrm{Re}\left[\dot{\mathcal{M}}^S(0)\right]},
\end{equation} 
in which we omit for simplicity the time-independent parametric arguments and the dot denotes the time derivative.\\

\end{document}